Spectral Properties and Bandstructure of Correlated Electron Systems

Th. Pruschke*, Th. Obermeier*, J. Keller* and M. Jarrell†

* Institut für Festkörperphysik, Universität Regensburg, 93040 Regensburg, Germany

† Department of Physics, University of Cincinnati, Cincinnati, OH 45221, USA

We present $\vec{k}$-dependent one-particle spectra and corresponding effective bandstructures for the $2d$ Hubbard model calculated within the dynamical molecular field theory (DMFT). This method has proven to yield highly nontrivial results for a variety of quantities but the question remains open to what extent it is applicable to relevant physical situations.

To address this problem we compare our results for spectral functions to those obtained by QMC simulations. The good agreement supports our notion that the DMFT is indeed a sensible ansatz for correlated models even in to $d = 2$.

The observation that the local electron-electron interaction in models of strongly correlated electronsystems leads in the limit spatial dimensions $d \to \infty$ to a local one-particle self energy [1], i.e. the single-particle propagator $G(\vec{k}, z)$ has the form $G(\vec{k}, z) = \left[G^{(0)}(\vec{k}, z)^{-1} - \Sigma(z)\right]^{-1}$ with $G^{(0)}(\vec{k}, z)$ as the corresponding propagator of the noninteracting system, founded the so-called *dynamical molecular-field theory*: The lattice problem maps onto an impurity system coupled to an effective, self-consistently determined bath [2] and the lattice structure enters only via the free density of states (DOS) $\rho_0(\epsilon) = \sum_{\vec{k}} \delta(\epsilon - \epsilon_{\vec{k}})$. The apparent advantages of this approach are that one (i) works in the thermodynamic limit and (ii) treats at least the important local correlations correctly.

This method has been successfully applied to study various physical properties of the one-band Hubbard model (HM) [3] $H_{\text{Hub}} = \sum_{\vec{k}\sigma} \epsilon_{\vec{k}} c^\dagger_{\vec{k}\sigma} c_{\vec{k}\sigma} + U \sum_l n_{l\uparrow} n_{l\downarrow}$ in the limit $d \to \infty$ [4, 5]. However, the question remains to what extent the dynamical molecular-field theory (DMFT) is applicable to real physical situation, especially to finite dimensions $d = 2, 3$ [6].

A first comparison between magnetic properties of the HM in $d = 3$ and $d = \infty$ [4] already showed that the latter limit seems indeed to provide a quite reliable way to describe at least static quantities of the HM. With respect to e.g. transport properties it is however far more interesting to know how well the DMFT describes the *dynamical* properties of correlated models in finite dimensions. Unfortunately, there are presently no (numerical) exact results for the dynamics of the HM in $d = 3$ since even Quantum Monte Carlo methods (QMC) are restricted to far too small system sizes to allow for a reliable evaluation of dynamical quantities like one-particle spectra. Recently, however, significant progress has been made in obtaining one-particle spectra from QMC-data for the HM on a two dimensional square-lattice [7, 8]. This enables the first direct and quantitative comparison between dynamical results from a generalized mean-field theory and numerical exact methods in finite dimensions.

Let us start the discussion of our results obtained by solving the dynamical molecular-field equations with NCA-techniques [9] by noting that the DMFT is relatively insensitive to the actual lattice used. This can be seen from Fig. 1, where we compare calculations for dispersions [1]

$$\epsilon_{\vec{k}} = -\frac{t^*}{\sqrt{d}} \sum_{l=1}^{d} \cos(k_l a) \ . \tag{1}$$

with $d = 2$ (dashed curves) and $d = \infty$ (full curves). As is evident from the inset to Fig. 1a, the DOS for the noninteracting systems show remarkable differences, namely sharp cut-offs at the band edges $\omega/t^* = \pm\sqrt{2}$ and a van-Hove singularity at $\omega = 0$ for $d = 2$ in contrast to a featureless DOS in $d = \infty$ (Gaussian). Nevertheless the DOS for finite Coulomb energy $U = 4t^*$ shows no significant differences in both cases in Fig. 1a.

A comprehensive picture of the structures of the single-particle spectrum can be obtained by searching for the positions of the maxima in $\rho(\vec{k}, \omega) = -\Im m G(\vec{k}, \omega + i0^+)/\pi$ and plotting them as a measure for the effective bandstructure of the interacting system. To allow a comparison between $d = 2$ and $d = \infty$, we use in both cases as the dispersion equation (1) with $d = 2$, but insert for $\Sigma(z)$ in $G(\vec{k}, z)$ either the result from our calculation with the Gaussian DOS ($d = \infty$) or with the two dimensional DOS ($d = 2$). Again, apart from slight modifications no significant differences can be observed. We thus conclude that the observed features of the DMFT are generic and do not depend on the underlying lattice. The spectral function $\rho(\vec{k}, \omega)$ for the HM along the $\Gamma - M$ direction of the Brillouin zone of the square lattice is shown in Fig. 2 for two fillings $\langle n \rangle = 1$ (Fig. 2a) and $\langle n \rangle = 0.95$ (Fig. 2b). The value for the Coulomb repulsion was chosen as $U = 4t^*$, so that for $\langle n \rangle = 1$ the system is insulating. While at half filling one observes only two rather broad structures below (lower Hubbard band) and above (upper Hubbard band) the Fermi energy, a third peak appears in addition close to the Fermi level for finite filling. This peak becomes extremely sharp when it crosses the Fermi energy and constitutes the quasiparticle states in the system.

In Fig. 3 we finally compare our results with QMC simulations of the two dimensional HM [8]. For clarity of the presentation we restrict ourselves to the effective bandstructures obtained from the peaks in the spectral functions. Fig. 3a gives an impression on the situation for half filling and large $U$. The temperature was chosen such that the antiferromagnetic correlation length was much smaller than the system size. As well in the DMFT (full curves) as in the QMC (points) one observes the two broad features (upper and lower Hubbard band) below and above $\mu$ and a good agreement of the overall bandwidth. The striking difference between DMFT and QMC is that the bands each seem to be split in the QMC. We do think, however, that this additional splitting is an artifact due to the small system size ($8 \times 8$), i.e. finite-size effects will lead to several distinct peaks instead of one broadened feature. The situation becomes better for a filling $\langle n \rangle = 0.95$ in Fig. 3b. Again the results from DMFT and QMC are superimposed. Here a good agreement concerning both the shape and the width of the bands is observed. Note that the high-energy results of the QMC again show the tendency to produce spurious splittings of the lower and upper Hubbard bands. Especially interesting is the clear existence of a narrow band close to $\mu$. Such a band was already noted earlier in perturbation theories [10] and in the DMFT [4]. The QMC results definitely show that it is indeed a general feature of the HM in $d \geq 2$ and not an artifact of e.g. the DMFT.

To conclude we presented DMFT-results for spectral functions of the HM in two dimensions and compared our calculations with QMC simulations of the HM on a square lattice. As long as long-ranged spin correlations are unimportant, we find that the DMFT gives a rather accurate description of the dynamical properties of the HM even in $d = 2$. It is evident that the DMFT as a molecular-field theory is not able to properly describe the effects of long-ranged spin fluctuations on the dynamics of the system. We thus do expect and observe drastic deviations between DMFT and QMC as soon as the antiferromagnetic correlation length becomes of the order of the linear dimensions of the lattice [8]. Nevertheless we think that, as long as one keeps its limitations in mind, the DMFT is a valuable tool to study static and dynamic properties of strongly correlated electron systems and to supplement and extend results from other approaches.

This work was supported by the National Science Foundation grant number DMR-9406678 and DMR-9357199, the NATO Collaborative Research Grant number CRG 931429, the Deutsche Forschungsgemeinschaft grant number Pr 298/3-1 and through the NSF NYI program.

**Figure captions:**

**Fig. 1**: (a) Single-particle DOS for the Hubbard model calculated for a simple hypercubic lattice in $d = \infty$ (full line) and $d = 2$ (dashed line). The parameters are $U = 4t^*$, $T = t^*/30$ and $\langle n \rangle = 0.95$. The inset shows the corresponding DOS for the noninteracting system. (b) Effective bandstructure for a square lattice obtained from the peaks of $\rho(\vec{k}, \omega)$.

**Fig. 2**: Spectral function for the HM on a square lattice along the $\Gamma$-$M$ line of the Brillouin zone. (a) Half filling $\langle n \rangle = 1$ and (b) a filling $\langle n \rangle = 0.95$. Other parameters are $U = 4t^*$ and $T = t^*/30$.

**Fig. 3**: Comparison of the results for the effective bandstructure of the HM on a square lattice as obtained from DMFT (full curves) and QMC (points and shaded areas) [8]. (a)$\langle n \rangle = 1$, $U = 4t^*$ and $T = t^*/7$. (b) $\langle n \rangle = 0.95$, $U = 2.82t^*$ und $T = t^*/28.8$.

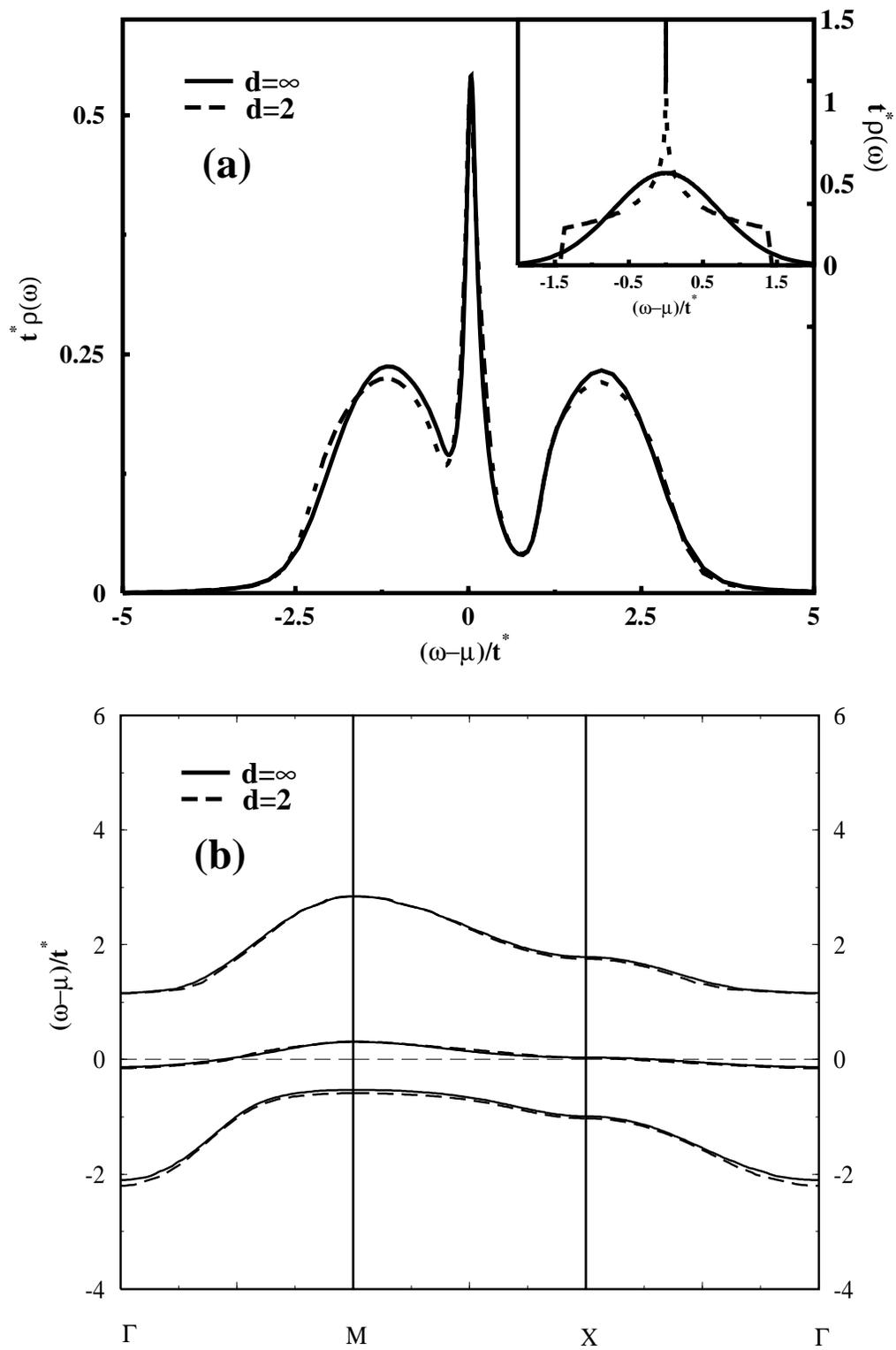

Figure 1

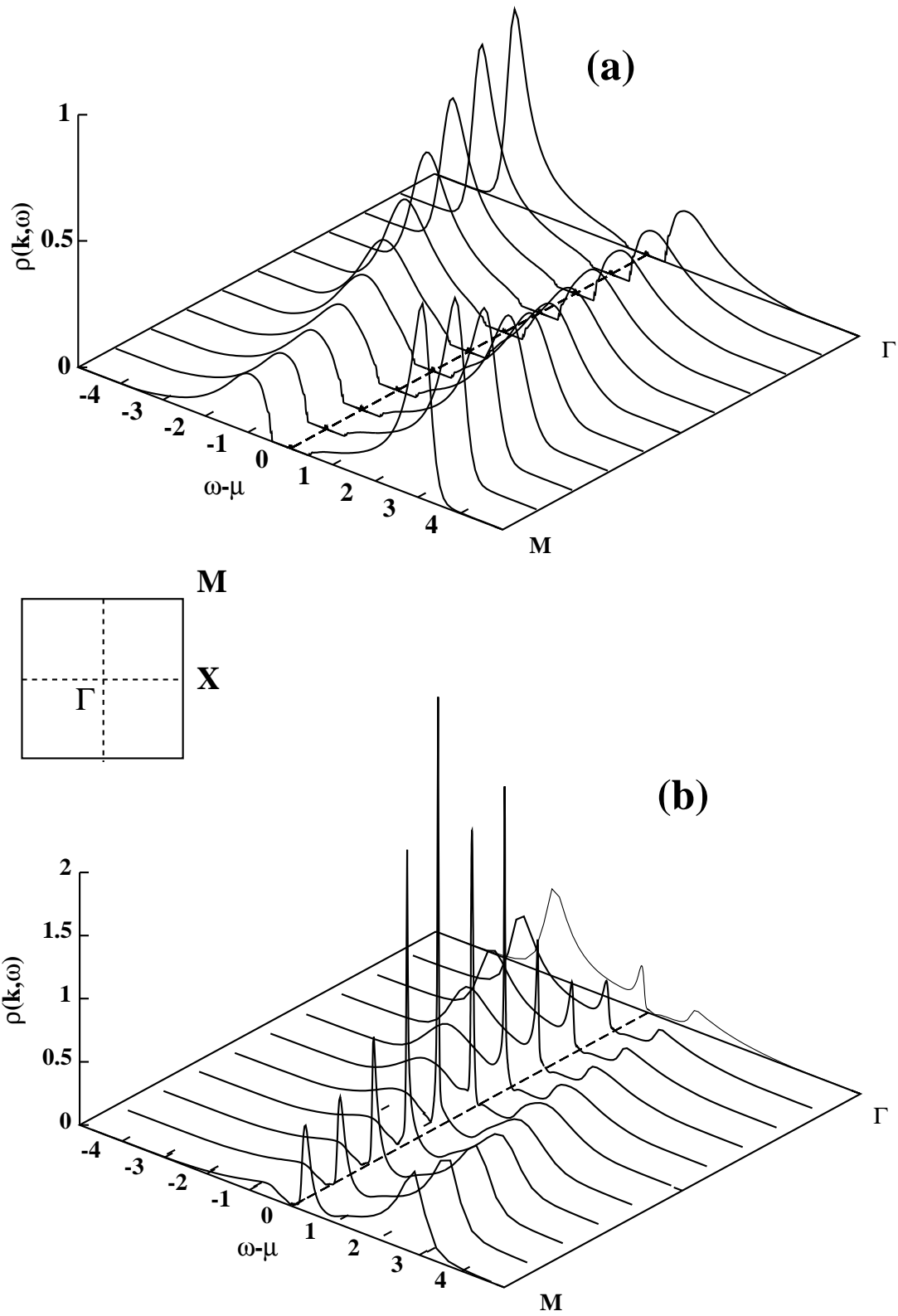

Figure 2

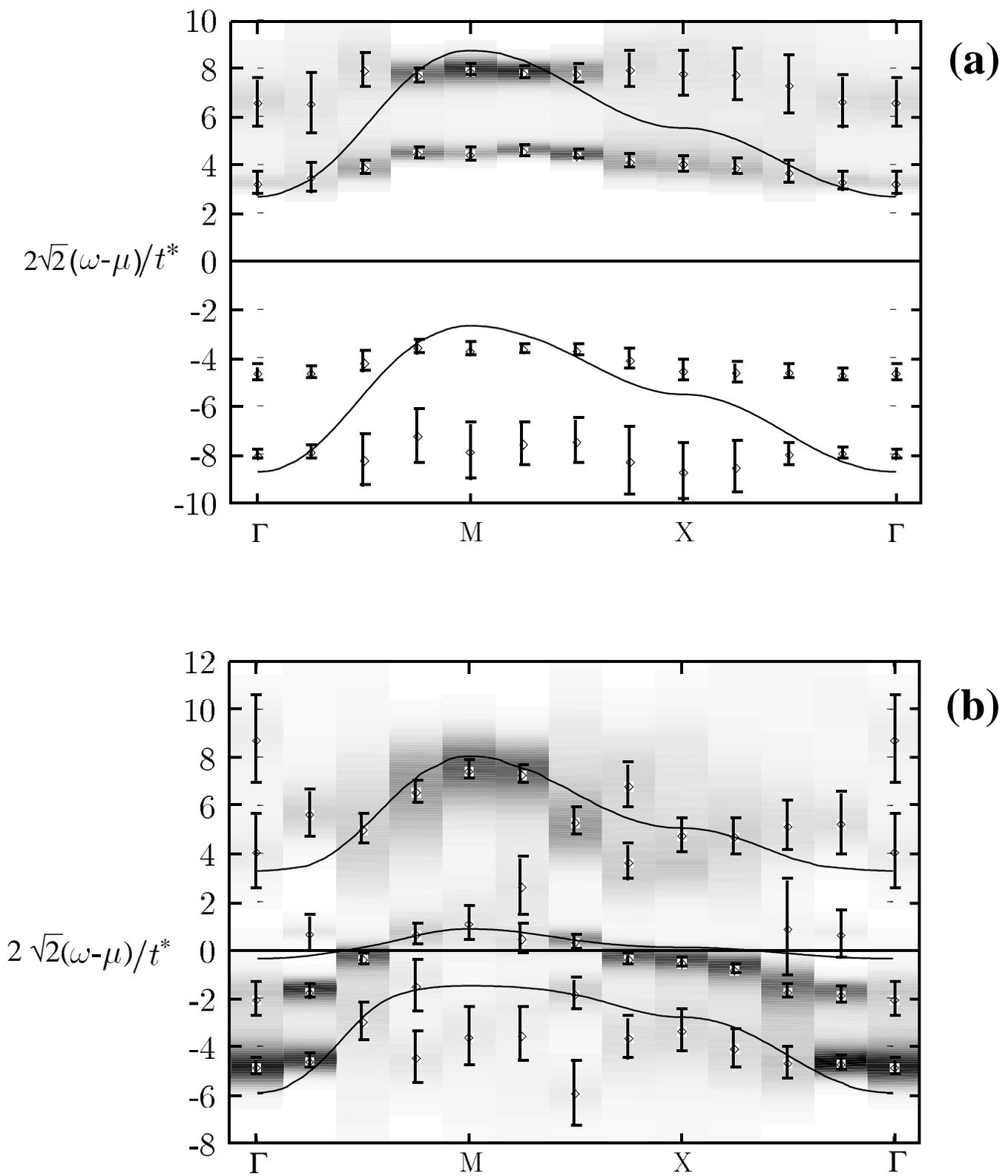

Figure 3